\documentclass[english]{article}
\usepackage[T1]{fontenc}
\usepackage[latin9]{inputenc}
\usepackage{verbatim}
\usepackage{float}
\usepackage{calc}
\usepackage{amsmath}
\usepackage{graphicx}
\usepackage{wasysym}
\usepackage{extarrows}
\usepackage{cite}

\usepackage{color}

\makeatletter

\providecommand{\tabularnewline}{\\}

\usepackage{amssymb}
\usepackage{latexsym}
\usepackage{epsfig}
\usepackage{float}
\usepackage{dsfont}

\newcommand{\be}{\begin{equation}}
\newcommand{\ee}{\end{equation}}

\makeatother

\usepackage{babel}
\begin{document}
{}~ \hfill\vbox{\hbox{CTP-SCU/2018007}}\break
\vskip 0.5cm
\centerline{\Large \bf  Fix the dual geometries of $T\bar{T}$ deformed CFT$_2$}
\vspace*{1.0ex}
\centerline{\Large \bf  and highly excited states of CFT$_2$}

\vspace*{3.0ex}
\centerline{\large Peng Wang, Houwen Wu and Haitang Yang}
\vspace*{3.0ex}
\centerline{\large \it Center for theoretical physics}
\centerline{\large \it Sichuan University}
\centerline{\large \it Chengdu, 610064, China} \vspace*{1.0ex}
\vspace*{3.0ex}

\centerline{pengw@scu.edu.cn, iverwu@scu.edu.cn, hyanga@scu.edu.cn}
\vspace*{3.0ex}
\centerline{\bf Abstract} \bigskip 
In previous works, we have developed an approach to fix the leading behaviors of the pure AdS$_3$ and BTZ black hole from the entanglement entropies of the free CFT$_2$ and  finite temperature CFT$_2$, respectively. We exclusively use holographic principle only and make no restriction about the bulk geometry, not only the kinematics but also the dynamics.  In order to verify the universality and correctness of our method, in this paper, we apply it to  the $T\bar{T}$ deformed CFT$_2$, which breaks the conformal symmetry. In terms of the physical arguments of the $T\bar{T}$ deformed CFT$_2$, the derived metric is a deformed BTZ black hole. The requirement that the CFT$_2$  lives on  a conformally flat boundary leads to  $r_{c}^{2}=\ 6R_{AdS}^{4}/(\pi c\mu)$ naturally,  in perfect agreement with previous conjectures in literature. The energy spectum and propagation speed calculated with this deformed BTZ metric are the same as these derived from $T\bar{T}$ deformed CFT$_2$. We furthermore fix the dual geometry of highly excited states  with our approach. The result contains the descriptions for the conical defect and BTZ black hole.

\baselineskip=14pt

\section{Introduction and summary}

The $T\bar{T}$ deformed CFT$_{2}$, discovered by the remarkable
works \cite{Zamolodchikov:2004ce,Cavaglia:2016oda,Smirnov:2016lqw},
attracts increasing attention recently. It is a solvable example of
quantum field theory. A conformal field theory (CFT), which is conformally
invariant at different fixed points along the RG flow, can be deformed
by relevant, irrelevant and marginal deformations. The irrelevant deformation
only affects the physics in the UV region
but not the IR region. $T\bar{T}$ deformation supports a class of
solvable irrelevant deformations of CFT$_{2}$ by turning on the $T\bar{T}$
term. The integrability implies the theory contains an infinite element
of conserved charges. The class of $T\bar{T}$ deformed CFT$_{2}$
is characterized by a length squared parameter $\mu\geq0$ as $dS^\mu/d\mu = \int d^2 x (T\bar T)_\mu$, where $(T\bar T)_\mu=\frac{1}{8}\left(T^{\alpha\beta}T_{\alpha\beta}-\left(T_{\alpha}^{\alpha}\right)^{2}\right)_\mu$ is defined in terms of the stress tensor   of the deformed theory. As $\mu\to 0$, the simplest example is

\begin{equation}
S^{\mu}=S_{\mathrm{CFT}}+\mu\int d^{2}x\,T\bar{T},
\end{equation}

\noindent  It is clear that $T\bar{T}$
deformation is Lorentz invariant and breaks the conformal symmetry.
McGough, Mezei, and Verlinde \cite{McGough:2016lol} proposed that the $T\bar{T}$ deformed
CFT$_{2}$ is no longer located on the asymptotic boundary of AdS$_{3}$
but lives at the finite radial position $r_{c}$\footnote{Note that to compare with \cite{McGough:2016lol}, our $\mu$ should be replaced by  $\mu/(4\pi^2)$  for notation difference.},

\begin{equation}
r_{c}^{2}=\frac{6R_{AdS}^{4}}{\pi c\mu},\label{eq:radial}
\end{equation}

\noindent where $R_{AdS}$ is the radius of AdS$_{3}$ and $c=\frac{3R_{AdS}}{2G}$
is the central charge of the CFT$_{2}$. Based on this picture, a
new way to extend and study the AdS/CFT correspondence emerges. In
a recent work \cite{Chen:2018eqk}, Chen and  collaborators calculated
the entanglement entropy of the $T\bar{T}$ deformed CFT$_{2}$ by
the replica trick. They found that for a finite size system, there
is no leading correction. But for a finite temperature system, they
obtained

\begin{equation}
S_{EE}^{\mu}=\frac{c}{3}\log\left(\frac{\beta}{\pi a}\sinh\left(\frac{\pi\triangle x}{\beta}\right)\right)-\mu\frac{\pi^{4}c^{2}\triangle x}{9\beta^{3}}\coth\left(\frac{\pi\triangle x}{\beta}\right).
\end{equation}

\noindent It is curious
why there is no correction to the finite size system. Without a correction,
how do we tell the CFT lives at a finite radial distance for a finite
size system by the entanglement entropy? We speculate the reason
is that the entanglement entropy is defined for spacelike intervals
only, which is not consistent with the Lorentz covariance. Since the
finite size and finite temperature systems have the same cylindrical
topology under exchanging $t\leftrightarrow x$, we are inspired to
speculate that there exists a temporal entanglement entropy for a
temporal interval (indicating entangling between the big bang and
big crunch?), and the finite size system receives a correction in
this temporal entanglement entropy. Unfortunately, up to now, there
is no well-defined temporal entanglement entropy. We are thus led
to believe, once a temporal entanglement entropy is correctly introduced,
one would find the results as shown in Table (\ref{tab:correction}):

\begin{table}[H]
\begin{centering}
\begin{tabular}{|c|c|c|}
\hline
 & space-like & time-like\tabularnewline
\hline
Finite size &  & $\checked$\tabularnewline
\hline
Finite temperature & $\checked$ & \tabularnewline
\hline
\end{tabular}
\par\end{centering}
\caption{\label{tab:correction}The symbol $\checked$ marks   a correction to
the entanglement entropy caused by the $T\bar{T}$ deformation. }
\end{table}

In the same paper, Chen etc. verified the holographic method (RT formula
\cite{Ryu:2006bv}) by calculating the geodesic length in the BTZ
background. To match the entanglement entropy of the $T\bar{T}$
deformed CFT$_{2}$, they found two endpoints of the geodesics must be fixed
on the finite cut-off $r_{c}$. The relevant paper \cite{Donnelly:2018bef} calculated entanglement entropy
of the $T\bar{T}$ deformed CFT$_{2}$ between two antipodal points on a sphere.
More recent works on the holographic studies of the
$T\bar{T}$ deformed CFT$_{2}$ can be found in the refs. \cite{Dubovsky:2012wk,Caselle:2013dra,Giveon:2017myj,Giveon:2017nie,Chakraborty:2018kpr,
Shyam:2017znq,Guica:2017lia,Giribet:2017imm,Dubovsky:2018bmo,Cottrell:2018skz,Bonelli:2018kik,Akhavan:2018wla}.

In previous works, we have developed a method to fix the leading behaviors of the dual geometries
from the entanglement entropies of the CFT$_{2}$ without using AdS/CFT
or imposing any constraints on the bulk geometry. It must be understood
that neither kinematics nor dynamics (Einstein equation) can be used
in the derivation, especially the latter one. Out of the general geometries,
gravity is those respecting the Einstein equation. Therefore, using
the Einstein equation in any place makes the derivations circular.
On the other hand, it is easy to understand that the symmetry argument
is a necessary but not a sufficient condition to determine the dual
geometry of a CFT. Furthermore, obviously, the $SO(2,2)$ symmetry
argument is no help for the $T\bar{T}$ deformed CFT$_{2}$ considered
in this paper, since this particular CFT breaks the $SO(2,2)$ symmetry.
Under these considerations, we successfully derived the asymptotic
AdS$_{3}$ from the infinitely long system of CFT$_{2}$ \cite{Wang:2017bym},
and 3D BTZ black hole from the finite temperature system of CFT$_{2}$
\cite{Wang:2018vbw}, both with time direction included as it should
be.

One of the distinct ingredients of our approach is to use Synge's
world function \cite{Synge:1960} to extract the spacetime metric from given geodesic
lengths:

\begin{equation}
g_{ij}=-\:\underset{x\rightarrow x'}{\lim}\partial_{x^{i}}\partial_{x^{j^{\prime}}}\left[\frac{1}{2}L^{2}\left(x,x^{\prime}\right)\right].\label{eq:derive metric}
\end{equation}

\noindent where $x^{i}=\left(x,y\right)$, and $L\left(x,x^{\prime}\right)$
is the length of the geodesic connecting two points $x^{i}$ and $x^{i'}$
in the unknown bulk geometry. It turns out that, in order to determine
the function form of the bulk geodesic length $L\left(x,x^{\prime}\right)$,
we need to use the entanglement entropies in UV and IR (mass gap)
regions. These entanglement entropies can fix the behaviors along
the boundary directions $x$, $t$ where the CFT lives, and the holographic
direction $y$ which is created by the energy cut-off of the CFT.
In order to verify the universality and correctness of our method,
we apply it to the $T\bar{T}$ deformed CFT$_{2}$ at finite temperature
to fix the leading behaviors of its dual geometry in this paper. The fixed metric turns
out to be a deformed BTZ metric in the planar coordinate, where ``deformed''
refers to a prefactor of $dx^{2}$ in contrast to the standard BTZ
metric:
\begin{eqnarray}
ds^{2} & = & \frac{R_{AdS}^{2}}{y^{2}}\left(-\left(1-\frac{y^{2}}{\beta_{H}^{2}}\right)dt^{2}+\left(1-\frac{\pi\mu c}{12\beta_{H}^{2}}+\ldots\right)^{2}dx^{2}+\left(1-\frac{y^{2}}{\beta_{H}^{2}}\right)^{-1}dy^{2}\right),\label{eq:deformed BTZ}\\
 & = & \frac{R_{AdS}^{2}}{y^{2}}\left(-\left(1-\frac{y^{2}}{\beta_{H}^{2}}\right)dt^{2}+d\tilde x^{2}+\left(1-\frac{y^{2}}{\beta_{H}^{2}}\right)^{-1}dy^{2}\right),\label{eq:BTZ}
\end{eqnarray}
\noindent where $\beta_H \equiv T^{-1}=\frac{\beta}{2\pi}$. It should
be emphasized that $x$ and $t$ in eqn. (\ref{eq:deformed BTZ})
are the physical (proper) arguments of the $T\bar{T}$ deformed
CFT$_{2}$. Though after a redefinition $\tilde x=(1-\frac{\pi\mu c}{12\beta_{H}^{2}}+\ldots)x$,
the metric (\ref{eq:BTZ}) is the standard BTZ, $\tilde x$ is the physical
argument of the original undeformed CFT. The advantage of the metric
(\ref{eq:deformed BTZ}) is that, once requiring the CFT lives on
the flat boundary of the dual geometry, we immediately deduce that
the geometry covers $y_{c}\leq y\leq \beta_{H}$, with $y_{c}^{2}=\frac{\pi\mu c}{6}$.
After transforming $y_{c}$ to the global coordinate, we get $r_{c}^{2}=\frac{6R_{AdS}^{4}}{\pi c\mu}$,
which is in perfect agreement with eqn. (\ref{eq:radial}). In
\cite{McGough:2016lol}, in order to check the duality, quantities
such as energy spectrum, thermodynamic properties and propagation
speeds were calculated based on the standard BTZ metric (\ref{eq:BTZ}),
and they were compared with the results from the $T\bar{T}$ deformed
CFT$_{2}$ to determine the value of $r_{c}$. Nevertheless, using
the deformed metric (\ref{eq:deformed BTZ}), in addition to determining $r_{c}^{2}=\frac{6R_{AdS}^{4}}{\pi c\mu}$  immediately, we straightforwardly
calculate the energy spectrum and propagation speeds, which are the
same as those from the $T\bar{T}$ deformed CFT$_{2}$. It is important to note that all these conclusions cannot be
obtained without the metric component $g_{tt}$.

As a further check, we next apply our approach to the highly excited
states of CFT$_{2}$ with large $c$ and sparse spectrum of low dimensional
operators. Using the entanglement entropies of such states of a finite size system, given in
\cite{Fitzpatrick:2014vua, Asplund:2014coa,Caputa:2014eta}, we fix
the dual geometry

\begin{equation}
ds^{2}=\frac{R_{AdS}^{2}}{y^{2}}\left(-\left(1-\frac{y^{2}}{\beta_{\Psi}^{2}}\right)dt^{2}+\left(1-\frac{y^{2}}{\beta_{\Psi}^{2}}\right)^{-1}\left(1+{\cal O}\left(\frac{y}{\beta_{\Psi}}\right)\right)dy^{2}+dx^{2}\right),
\end{equation}

\noindent where $\beta_{\Psi}=(\frac{12\triangle_{h}}{c}-1)^{-1/2}$.
When $\triangle_{h}<\frac{c}{12}$, the metric represents a conical
defect in the center of the global AdS$_{3}$. On the other hand,
when $\triangle_{h}>\frac{c}{12}$, the inserted operator is heavy
enough to produce a BTZ black hole at temperature $T=\beta_{\Psi}^{-1}$.

In literature, there is an alternative method to extract the bulk geometries from geodesics, namely, \emph{kinematic space}. It shows that the geodesics of $\mathrm{AdS}_{3}$ can be viewed as   points in the kinematic space by calculating the Crofton form from the entanglement entropy. The Crofton form is defined as the second derivatives with respect to two different points of the  entanglement entropy (not taking the limits afterwards!).
Then the points-grouped one-dimensional object, say point curve (codimension-1 locus), in the kinematic space corresponds to a set of geodesics which intersect with each other at a single point of $\mathrm{AdS}_{3}$. Therefore, geodesic length between any
two points in the bulk of $\mathrm{AdS}_{3}$ equals the integral
of Crofton form over the area between two corresponding point curves
in the kinematic space \cite{Czech:2015qta}. The advantage of kinematic space is that the correspondence between Crofton\textquoteright s formula and field theory is clear. Then, it is possible to obtain the OPE blocks from the local operators in the AdS by Radon transformation \cite{Czech:2016xec,deBoer:2016pqk}. However, it is  not easy and straightforward to fix the leading behaviors of the spacetime metric of previously unknown bulk geometry. One of the main differences and advantages of our approach is to impose the limit $x\to x'$ after taking the derivatives, which enables us to get the metric  immediately from the geodesics, as shown in eqn.(\ref{eq:derive metric}).

The reminder of this paper is outlined as follows. In section 2, we
show how to fix the leading behaviors of the deformed BTZ spacetime from the entanglement
entropies of the $T\bar{T}$ deformed CFT$_{2}$, and calculate the energy spectrum and propagation speeds
with the deformed metric. In section 3, we
address the highly excited states of CFT$_{2}$ and fix the leading behaviors of its dual
geometry from the entanglement entropy. Section 4 is for conclusion.

\section{The dual geometry of the $T\bar{T}$ deformed CFT$_2$}

To the leading order correction, the UV entanglement entropy of $T\bar{T}$
deformed CFT$_{2}$ at finite temperature was calculated in \cite{Chen:2018eqk},

\begin{equation}
S_{EE}^{\mu}=\frac{c}{3}\log\left(\frac{2\beta_{H}}{a}\sinh\left(\frac{\triangle x}{2\beta_{H}}\right)\right)-\pi\mu\frac{c^{2}\triangle x}{72\beta_{H}^{3}}\coth\left(\frac{\triangle x}{2\beta_{H}}\right),\label{eq:Corrected EE}
\end{equation}

\noindent where $\beta_{H}\equiv T^{-1}=\frac{\beta}{2\pi}$ and $a$
is a UV cut-off. As $\mu/\beta_{H}^{2}\rightarrow0$, it reduces to
the entanglement entropy of the undeformed CFT$_{2}$ at finite temperature:

\begin{equation}
S_{EE}=\frac{c}{3}\log\left(\frac{2\beta_{H}}{a}\sinh\left(\frac{\triangle x}{2\beta_{H}}\right)\right).\label{eq:finite EE}
\end{equation}

\noindent Since the correction is a function of $\triangle x/\beta_{H}$
and $\mu/\beta_{H}^{2}$, it is tempting to check if it can be captured by
an effective correction to $\triangle x$. This step is consistent
with the fact that irrelevant deformations only affect the physics
in the UV region. Therefore, we set

\begin{equation}
S_{EE}^{\mu}=\frac{c}{3}\log\left(\frac{2\beta_{H}}{a}\sinh\left(\frac{\triangle x}{2\beta_{H}}\left(1+F\left(\mu,\triangle x\right)\right)\right)\right),
\end{equation}

\noindent where $F\left(\mu,\triangle x\right)$ is a correction to
$\triangle x$. It is remarkable to notice for small $F(\mu,\Delta x)$,
the Taylor expansion is

\begin{equation}
S_{EE}^{\mu}=\frac{c}{3}\log\left(\frac{2\beta_{H}}{a}\sinh\left(\frac{\triangle x}{2\beta_{H}}\right)\right)+F\left(\mu,\triangle x\right)\frac{c\triangle x}{6\beta_{H}}\coth\left(\frac{\triangle x}{2\beta_{H}}\right)+\ldots.
\end{equation}

\noindent Therefore, comparing with eqn. (\ref{eq:Corrected EE}),
we can get

\begin{equation}
F\left(\mu,\triangle x\right)=-\frac{\pi\mu c}{12\beta_{H}^{2}}+\mathcal{O}(\mu^{2})
\end{equation}

\noindent So, the UV entanglement entropy of $T\bar{T}$ deformed
CFT$_{2}$ can be written as:

\begin{equation}
S_{EE}^{\mu}=\frac{c}{3}\log\left(\frac{2\beta_{H}}{a}\sinh\left(\frac{\triangle x}{2\beta_{H}}\left(1-\frac{\pi\mu c}{12\beta_{H}^{2}}+\ldots\right)\right)\right).\label{eq:TT EE}
\end{equation}

\noindent The time-dependent entanglement entropy of the undeformed
CFT$_{2}$ at finite temperature is:

\begin{equation}
S_{EE}\left(t\right)=\frac{c}{3}\log\left(\frac{\beta_{H}}{a}\left[\sqrt{2\cosh\left(\frac{\triangle x}{\beta_{H}}\right)-2\cosh\left(\frac{\triangle t}{\beta_{H}}\right)}\right]\right).\label{eq:covar EE}
\end{equation}

\noindent As $\triangle t=0$, it reduces to eqn. (\ref{eq:finite EE}).
For the $T\bar{T}$ deformation, we only need to replace $\triangle x$
by $\triangle x\left(1-\frac{\pi\mu c}{12\beta_{H}^{2}}+\ldots\right)$
to get the $T\bar{T}$ deformed one

\begin{equation}
S_{EE}^{\mu}\left(t\right)=\frac{c}{3}\log\left(\frac{\beta_{H}}{a}\left[\sqrt{2\cosh\left(\frac{\triangle x}{\beta_{H}}\left(1-\frac{\pi\mu c}{12\beta_{H}^{2}}+\ldots\right)\right)-2\cosh\left(\frac{\triangle t}{\beta_{H}}\right)}\right]\right).
\end{equation}

\noindent Setting $R\equiv\frac{2G_{N}^{\left(3\right)}c}{3}$ and
identifying this entanglement entropy with the length of the geodesic
anchored on the boundary in the dual geometry, one has

\begin{eqnarray}
\frac{L_{\mathrm{boundary}}}{R} & = & \log\left(\frac{2\beta_{H}^{2}}{a^{2}}\left[\cosh\left(\frac{\triangle x}{\beta_{H}}\left(1-\frac{\pi\mu c}{12\beta_{H}^{2}}+\ldots\right)\right)-\cosh\left(\frac{\triangle t}{\beta_{H}}\right)\right]\right).\label{eq:BTZ boundary L}
\end{eqnarray}

\noindent From the holographic principle, the energy cut-off $a$
generates a holographic dimension $y$, say $a^{2}\rightarrow yy^{\prime}\left(1+\ldots\right)$.
In order to apply eqn. (\ref{eq:derive metric}) to get the metric,
this ending on boundary geodesic length eqn. (\ref{eq:BTZ boundary L})
is obviously not sufficient and we need to push the endpoints into
the bulk, with dependence on the holographic dimension $y$, to describe
bulk geodesic connecting arbitrary endpoints $(t,x,y)$ and $(t',x',y')$.
On the other hand, as shown by eqn. (\ref{eq:Corrected EE}), when
$\beta_{H}\rightarrow\infty$, the entanglement entropy becomes the free CFT$_2$ one,  whose dual geometry is the pure AdS$_{3}$ in Poincare patch, as we derived in \cite{Wang:2017bym} by our approach. The geodesic length of pure AdS$_{3}$ in Poincare patch  connecting $(t,x,y)$ and $(t',x',y')$ is

\begin{equation}
\cosh\left(\frac{L_{\mathrm{bulk}}}{R}\right)=\frac{\left(\triangle x\right)^{2}-\left(\triangle t\right)^{2}+y^{2}+y^{\prime2}}{2yy^{\prime}}.\label{eq:pure AdS L}
\end{equation}

\noindent To meet these two requirements (\ref{eq:BTZ boundary L})
and (\ref{eq:pure AdS L}) under different limits, we only have one
possibility to generalize the boundary anchored expression (\ref{eq:BTZ boundary L})
to the general bulk geodesic length:

\begin{eqnarray}
\cosh\left(\frac{L_{\mathrm{bulk}}}{R}\right) & = & \frac{\beta_{H}^{2}}{yy^{\prime}}\left[f\left(x,x',\frac{\mu}{\beta_{H}^{2}};y,y^{\prime};t,t'\right)\cosh\left(\frac{\triangle x}{\beta_{H}}\left(1-\frac{\mu c}{12\beta_{H}^{2}}+\ldots\right)\right)\right.\nonumber \\
 &  & \left.-g\left(x,x',\frac{\mu}{\beta_{H}^{2}};y,y^{\prime};t,t'\right)\cosh\left(\frac{\triangle t}{\beta_{H}}\right)\right].\label{eq:BTZ cosh}
\end{eqnarray}

\noindent where $f\left(x,x^{\prime},\frac{\mu}{\beta_{H}^{2}};y,y^{\prime};t,t'\right)$
and $g\left(x,x',\frac{\mu}{\beta_{H}^{2}};y,y^{\prime};t,t'\right)$
are arbitrary regular functions to keep the generalization universal.
The next steps are determining the behaviors of $f\left(x,x',\frac{\mu}{\beta_{H}^{2}};y,y^{\prime};t,t'\right)$
and $g\left(x,x',\frac{\mu}{\beta_{H}^{2}};y,y^{\prime};t,t'\right)$,
and then we can apply eqn. (\ref{eq:derive metric}) to get the metric.

\vspace{1em}

\noindent \textbf{Step 1:} When $\beta_{H}\gg y=y^{\prime}=a$ (geodesics
anchored on the boundary), $L_{\mathrm{bulk}}$
in eqn. (\ref{eq:BTZ cosh}) must reduce to $L_{\mathrm{boundary}}$,
given by equation (\ref{eq:BTZ boundary L}),
\begin{eqnarray}
L_{\mathrm{bulk}} & = & R\log\left(\frac{2\beta_{H}^{2}}{yy^{\prime}}\left[f\left(x,x',\frac{\mu}{\beta_{H}^{2}};y,y^{\prime};t,t'\right)\cosh\left(\frac{\triangle x}{\beta_{H}}\left(1-\frac{\mu c}{12\beta_{H}^{2}}+\ldots\right)\right)\right.\right.\nonumber \\
 &  & \left.\left.-g\left(x,x',\frac{\mu}{\beta_{H}^{2}};y,y^{\prime};t,t'\right)\cosh\left(\frac{\triangle t}{\beta_{H}}\right)\right]\right)\\
 & \rightarrow & R\log\left(\frac{2\beta_{H}^{2}}{a^{2}}\left[\cosh\left(\frac{\triangle x}{\beta_{H}}\right)-\cosh\left(\frac{\triangle t}{\beta_{H}}\right)\right]\right).
\end{eqnarray}
\noindent So as $\beta_{H}\gg y$ and $y^{\prime}$, we have
\begin{eqnarray}
f\left(x,x',\frac{\mu}{\beta_{H}^{2}};y,y^{\prime};t,t'\right) & = & 1+\varrho_{1}\left(x,x',\frac{\mu}{\beta_{H}^{2}};t,t'\right)\left(\frac{y}{\beta_{H}}+\frac{y^{\prime}}{\beta_{H}}\right)+\varrho_{2}\left(x,x',\frac{\mu}{\beta_{H}^{2}};t,t'\right)\left(\frac{y}{\beta_{H}}+\frac{y^{\prime}}{\beta_{H}}\right)^{2}+\ldots\nonumber \\
 &  & +\rho_{1}\left(x,x',\frac{\mu}{\beta_{H}^{2}};t,t'\right)\left(\frac{yy^{\prime}}{\beta_{H}^{2}}\right)+\rho_{2}\left(x,x',\frac{\mu}{\beta_{H}^{2}};t,t'\right)\left(\frac{yy^{\prime}}{\beta_{H}^{2}}\right)^{2}+\ldots,\nonumber \\
\nonumber \\
g\left(x,x',\frac{\mu}{\beta_{H}^{2}};y,y^{\prime};t,t'\right) & = & 1+\bar{\varrho}_{1}\left(x,x',\frac{\mu}{\beta_{H}^{2}};t,t'\right)\left(\frac{y}{\beta_{H}}+\frac{y^{\prime}}{\beta_{H}}\right)+\bar{\varrho}_{2}\left(x,x',\frac{\mu}{\beta_{H}^{2}};t,t'\right)\left(\frac{y}{\beta_{H}}+\frac{y^{\prime}}{\beta_{H}}\right)^{2}+\ldots\nonumber \\
 &  & +\bar{\rho}_{1}\left(x,x',\frac{\mu}{\beta_{H}^{2}};t,t'\right)\left(\frac{yy^{\prime}}{\beta_{H}^{2}}\right)+\bar{\rho}_{2}\left(x,x',\frac{\mu}{\beta_{H}^{2}};t,t'\right)\left(\frac{yy^{\prime}}{\beta_{H}^{2}}\right)^{2}+\ldots,
\end{eqnarray}
\noindent where $\varrho_{i}$, $\rho_{i}$, $\bar{\varrho}_{i}$
and $\bar{\rho}_{i}$ are regular and bounded functions.

\vspace{1em}

\noindent \textbf{Step 2:} As $\beta_{H}\rightarrow\infty$, or $\mu/\beta_{H}^{2}\ll1$
and $\beta_{H}\gg\triangle x$, $\triangle t$, $y$ and $y^{\prime}$,
the general expression (\ref{eq:BTZ cosh}) must match the pure AdS$_{3}$
background (\ref{eq:pure AdS L}).  From step 1, we know the leading term of $f$ and $g$ is the unit.
So, we have
\begin{eqnarray}
 &  & \cosh\left(\frac{L_{\mathrm{bulk}}}{R}\right)\nonumber \\
 & \simeq & \frac{\beta_{H}^{2}}{yy^{\prime}}\left[f\left(x,x^{\prime},\frac{\mu}{\beta_{H}^{2}};y,y^{\prime};t,t'\right)\left(1+\frac{\left(\triangle x\right)^{2}}{2\beta_{H}^{2}}-\frac{\pi\mu c\triangle x}{12\beta_{H}^{3}}+\ldots\right)-g\left(x,x^{\prime},\frac{\mu}{\beta_{H}^{2}};y,y^{\prime};t,t'\right)\left(1+\frac{\left(\triangle t\right)^{2}}{2\beta_{H}^{2}}+\ldots\right)\right]\nonumber \\
 & = & \frac{1}{2yy^{\prime}}\left[f\left(x,x^{\prime},\frac{\mu}{\beta_{H}^{2}};y,y^{\prime};t,t'\right)\left(\triangle x\right)^{2}-g\left(x,x^{\prime},\frac{\mu}{\beta_{H}^{2}};y,y^{\prime};t,t'\right)\left(\triangle t\right)^{2}\right.\nonumber \\
 &  & \left.+2\beta_{H}^{2}\left(f\left(x,x^{\prime},\frac{\mu}{\beta_{H}^{2}};y,y^{\prime};t,t'\right)-g\left(x,x^{\prime},\frac{\mu}{\beta_{H}^{2}};y,y^{\prime};t,t'\right)\right)+\ldots\right],\label{eq:temp cond.3} \\
 & \rightarrow & \frac{\left(\triangle x\right)^{2}-\left(\triangle t\right)^{2}+y^{2}+y^{\prime2}}{2yy^{\prime}},\label{eq:cond.3}
\end{eqnarray}
\noindent where we only keep the leading order. When keeping the higher
orders, we suppose to get the metric of asymptotic
AdS by using equation (\ref{eq:derive metric}). Or equivalently speaking,
since $f\left(x,x^{\prime},\frac{\mu}{\beta_{H}^{2}};y,y^{\prime};t,t'\right)$
and $g\left(x,x^{\prime},\frac{\mu}{\beta_{H}^{2}};y,y^{\prime};t,t'\right)$
lead to the coefficients of $dt^{2}$ and $dx^{2}$, to be an
asymptotic AdS, the regular functions $f\left(x,x^{\prime},\frac{\mu}{\beta_{H}^{2}};y,y^{\prime};t,t'\right)$
and $g\left(x,x^{\prime},\frac{\mu}{\beta_{H}^{2}};y,y^{\prime};t,t'\right)$
can only depend on $y$ and $y'$. Thus the general expression of
the bulk geodesic length takes the form


\noindent\fbox{\begin{minipage}[t]{1\columnwidth - 2\fboxsep - 2\fboxrule}%
\begin{equation}
\cosh\left(\frac{L_{\mathrm{bulk}}}{R}\right)=\frac{\beta_{H}^{2}}{yy^{\prime}}\left[f\left(\frac{y}{\beta_{H}},\frac{y'}{\beta_{H}}\right)\cosh\left(\frac{\triangle x}{\beta_{H}}\left(1-\frac{\pi\mu c}{12\beta_{H}^{2}}+\ldots\right)\right)-g\left(\frac{y}{\beta_{H}},\frac{y'}{\beta_{H}}\right)\cosh\left(\frac{\triangle t}{\beta_{H}}\right)\right],\label{eq:general expression of L}
\end{equation}
\end{minipage}}
\vspace{1em}

\noindent where
\begin{eqnarray}
f\left(\frac{y}{\beta_{H}},\frac{y^{\prime}}{\beta_{H}}\right) & = & 1+\varrho_{1}\left(\frac{y}{\beta_{H}}+\frac{y^{\prime}}{\beta_{H}}\right)+\varrho_{2}\left(\frac{y}{\beta_{H}}+\frac{y^{\prime}}{\beta_{H}}\right)^{2}+\ldots\nonumber \\
 &  & +\rho_{1}\left(\frac{yy^{\prime}}{\beta_{H}^{2}}\right)+\rho_{2}\left(\frac{yy^{\prime}}{\beta_{H}^{2}}\right)^{2}+\ldots,\nonumber \\
\nonumber \\
g\left(\frac{y}{\beta_{H}},\frac{y^{\prime}}{\beta_{H}}\right) & = & 1+\bar{\varrho}_{1}\left(\frac{y}{\beta_{H}}+\frac{y^{\prime}}{\beta_{H}}\right)+\bar{\varrho}_{2}\left(\frac{y}{\beta_{H}}+\frac{y^{\prime}}{\beta_{H}}\right)^{2}+\ldots\nonumber \\
 &  & +\bar{\rho}_{1}\left(\frac{yy^{\prime}}{\beta_{H}^{2}}\right)+\bar{\rho}_{2}\left(\frac{yy^{\prime}}{\beta_{H}^{2}}\right)^{2}+\ldots
\end{eqnarray}
\noindent and
\begin{equation}
f\left(\frac{y}{\beta_{H}},\frac{y^{\prime}}{\beta_{H}}\right)-g\left(\frac{y}{\beta_{H}},\frac{y^{\prime}}{\beta_{H}}\right)=\frac{1}{2\beta_{H}^{2}}\left(y^{2}+y^{\prime2}\right)+\mathcal{O}\left(\frac{1}{\beta_{H}^{4}}\right),
\end{equation}

\noindent given by the last two terms of (\ref{eq:temp cond.3}).

\vspace{1em}

\noindent \emph{Since the correction $\frac{\triangle x}{\beta_{H}}\left(1-\frac{\mu c}{12\beta_{H}^{2}}+\ldots\right)$
appears in the $\cosh$-function only, it does not affect determining
the behaviors of functions $f\left(\frac{y}{\beta_{H}},\frac{y^{\prime}}{\beta_{H}}\right)$
and $g\left(\frac{y}{\beta_{H}},\frac{y^{\prime}}{\beta_{H}}\right)$,
we are free to set $\mu=0$ in the following steps and the derivations
are same as what were done in our previous work \cite{Wang:2018vbw} for the case
of finite temperature CFT. After fixing  $f\left(\frac{y}{\beta_{H}},\frac{y^{\prime}}{\beta_{H}}\right)$
and $g\left(\frac{y}{\beta_{H}},\frac{y^{\prime}}{\beta_{H}}\right)$, we restore $\mu\not=0$}.

\vspace{1em}

\noindent \textbf{Step 3:}   When two endpoints of a geodesic coincide,
the geodesic length vanishes exactly. Plugging $x=x^{\prime}$, $y=y^{\prime}$
and $t=t^{\prime}$ into eqn. (\ref{eq:general expression of L}),
we get

\begin{equation}
f\left(\frac{y}{\beta_{H}},\frac{y}{\beta_{H}}\right)-g\left(\frac{y}{\beta_{H}},\frac{y}{\beta_{H}}\right)=\frac{y^{2}}{\beta_{H}^{2}}.
\label{eq:R5}
\end{equation}

\vspace{1em}

\noindent \textbf{Step 4:}
In ref. \cite{Takayanagi:2011zk}, T. Takayanagi proposed a new holographic dual of the boundary conformal field theory (BCFT). It presents that the phase transitions of entanglement entropy relate to the topological change of the RT surface in the bulk. Based on this realization, the boundary of BCFT will extend into the bulk and play a role in the end-of-the-world (ETW) brane \cite{Sully:2020pza}. The brane's tension corresponds to the boundary entropy of BCFT, and the RT surface which is anchored between the ETW in the bulk and the BCFT on the boundary relates to the entanglement entropy of BCFT. The region enclosed by ETW brane in the bulk and BCFT on the boundary is asymptotically AdS. Therefore, the dual RT surfaces of BCFT can be simply calculated between one point in the bulk and another point on the boundary in the AdS background without placing any new configuration, such as virtual branes which modify the bulk geometry \cite{Takayanagi:2011zk}. We are going to use this conclusion in this step,  since our method only depends on the RT surfaces and aims to fix the leading behaviors of bulk geometry.

\begin{figure}[H]
\begin{centering}
\includegraphics[scale=0.3]{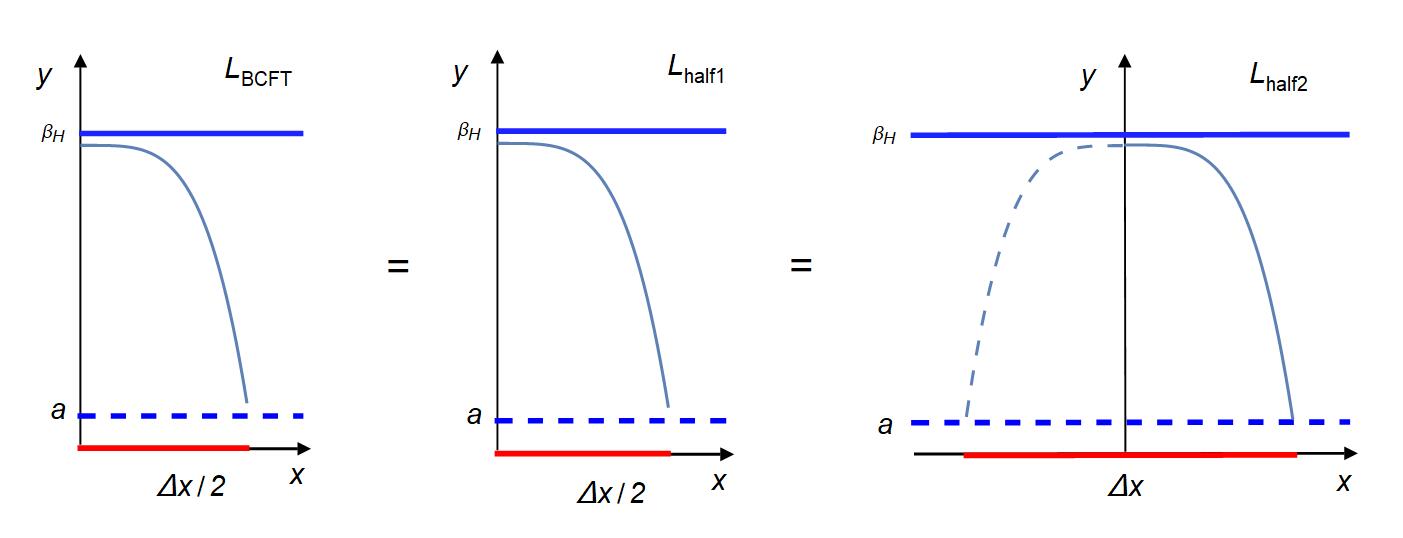}
\caption{The left picture is given by BCFT with $\Delta x\to\infty$. The middle picture is obtained
from $L_{\mathrm{bulk}}$ by setting $y=a$, $y^{\prime}=\beta_{H}$
and $\triangle x/2\rightarrow\infty$. The right picture is also given
by $L_{\mathrm{bulk}}$ from a different point of view, by setting
$y=y^{\prime}=a$ and $\Delta x\to\infty$. The solid lines in all
three pictures describe the same object. \label{fig:Three pics}}
\end{centering}
\end{figure}

We now consider the length of the segment in Fig. (\ref{fig:Three pics}) as $\Delta x\to\infty$. There are three ways to calculate it. The left picture is given in  \cite{Calabrese:2004eu,Calabrese:2009qy} from  boundary conformal field theory (BCFT),
\begin{equation}
L_{\mathrm{BCFT}}=R\log\left[\frac{\beta_{H}}{a}\exp\left(\frac{\triangle x}{2\beta_{H}}\right)\right].\label{eq:half 1}
\end{equation}

\noindent On the other hand, by using eqn. (\ref{eq:general expression of L}), we have two other
ways to calculate it. The first way is to
straightforwardly substitute $y=a$, $y'=\beta_{H}$, $\Delta x/2\to\infty$
into (\ref{eq:general expression of L}) to get

\begin{equation}
L_{\mathrm{bulk}}\rightarrow L_{\mathrm{half1}}=R\log\left(\frac{\beta_{H}}{a}\,f\left(\frac{a}{\beta_{H}},\frac{\beta_{H}}{\beta_{H}}\right)\exp\left(\frac{\triangle x}{2\beta_H}\right)\right).\label{eq:half 2}
\end{equation}

\noindent It is easy to understand that this length is one half of
that connecting $y=y^{\prime}=a$, $\Delta x\to\infty$.
So the second way is

\begin{equation}
L_{\mathrm{bulk}}\rightarrow L_{\mathrm{half2}}=\frac{1}{2}\,R\log\left(\frac{\beta_{H}^{2}}{a^{2}}\,f\left(\frac{a}{\beta_{H}},\frac{a}{\beta_{H}}\right)\exp\left(\frac{\triangle x}{\beta_{H}}\right)\right).\label{eq:half 3}
\end{equation}

\noindent These three lengths (\ref{eq:half 1}), (\ref{eq:half 2})
and (\ref{eq:half 3}) ought to be identical. We thus obtain $f\left(\frac{a}{\beta_{H}},\frac{\beta_{H}}{\beta_{H}}\right)^{2}=f\left(\frac{a}{\beta_{H}},\frac{a}{\beta_{H}}\right)=1$. The derivation of this constraint does not require $\beta_{H}\gg a$.
As long as $\beta_{H}$ is the upper bound of $y$, the derivation
is justified. Since $a$ is a varying cut-off not beyond $\beta_{H}$,
satisfying $0<a/\beta_{H}\le1,$ we can safely replace $\frac{a}{\beta_{H}}$
by $\frac{y}{\beta_{H}}$ to get:

\begin{equation}
f\left(\frac{y}{\beta_{H}},\frac{\beta_H}{\beta_{H}}\right)^{2}=f\left(\frac{y}{\beta_{H}},\frac{y}{\beta_{H}}\right)=1.
\label{eq:R4}
\end{equation}

\vspace{1em}


\noindent \textbf{Step 5:} An important lesson we learned from the
free CFT$_{2}$ case in \cite{Wang:2017bym} is that, in order to completely determine the
dual geometry, we need to know the geodesic length between $a$ and
$\beta_{H}$ with $\triangle x=0$, i.e. the vertical geodesic. To
be consistent, this particular geodesic length must be provided by
the CFT$_{2}$ entanglement entropy. In the free CFT$_{2}$, the IR
entanglement entropy precisely fits the requirement.

Remarkably, we know that the finite
temperature CFT$_{2}$ and the finite size CFT$_{2}$ have the same geometry ${\mathbb R}\times {\mathbf S}^1$. We can either interpret it as a CFT on a compact spatial interval of size $L_S$, or view it as a thermal CFT on the real line with the Euclidean time along the circle with the period  $\beta = L_S$, as explained in \cite{Calabrese:2004eu, Rangamani:2016dms} in detail. So these two CFTs are basically the same scenario and  have the same bulk dual. We are allowed to use the results from both CFTs to construct the dual bulk geometry with the identification

\begin{equation}
\beta_{H}=\frac{\beta}{2\pi}\leftrightarrow\frac{L_{S}}{2\pi}.
\end{equation}

\begin{figure}[H]
\begin{centering}
\includegraphics[scale=0.4]{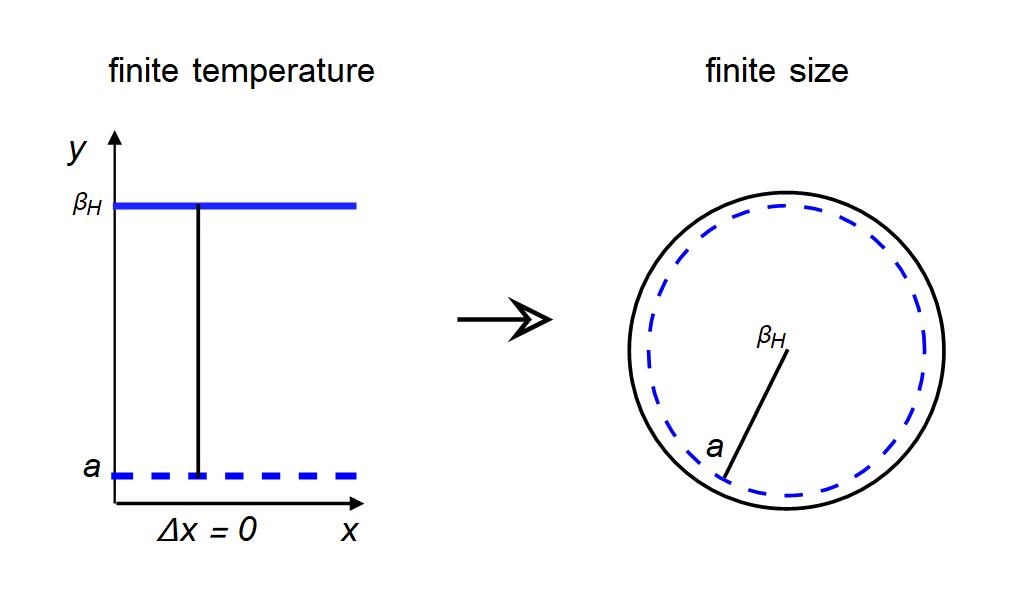}
\par\end{centering}
\centering{}\caption{\label{fig:The-geodesic-between}The geodesic between $y=a$ and $y^{\prime}=\beta_{H}$
at a finite temperature system is mapped to a finite size system,
corresponding to the radius of a circle from $y=a$ to $y=\frac{L_S}{2\pi}$.}
\end{figure}

\noindent Therefore, the geodesic length between $a$ and $\beta_{H}=\frac{\beta}{2\pi}$
in the finite temperature system can be obtained from the geodesic length connects $a$ and $\frac{L}{2\pi}$
in the finite size system,
\begin{eqnarray}
\mathrm{finite\;temperature} &  & \mathrm{finite\;size}\nonumber \\
L_{\mathrm{geodesic}}\left(a,\frac{\beta}{2\pi}\right) & \xlongequal{\beta\leftrightarrow L_S} & L_{\mathrm{geodesic}}\left(a,\frac{L_{S}}{2\pi}\right).
\end{eqnarray}
\noindent Noting that $\frac{L_{S}}{2\pi}$ is the radius of the finite
size system with the circumference $L_{S}$, therefore, this geodesic
goes from boundary to the center of the circle, as illustrated in
Fig. (\ref{fig:The-geodesic-between}).

\noindent We know that the entanglement entropy of a finite size system is\footnote{According to \cite{Chen:2018eqk}, the entanglement entropy of a finite size system receives no correction.}

\begin{equation}
S_{EE}=\frac{c}{3}\log\left(\frac{L_{S}}{\pi a}\sin\left(\frac{\pi\triangle x}{L_{S}}\right)\right),
\end{equation}

\noindent The maximal entanglement entropy is achieved by splitting
the circle into two equal regions, $\Delta x=L_{S}/2$. The corresponding
geodesic is nothing but a diameter

\begin{equation}
S_{EE}=\frac{c}{3}\log\left(\frac{L_{S}}{\pi a}\right),\qquad L_{\mathrm{boundary}}=2R\log\left(\frac{L_{S}}{\pi a}\right).
\end{equation}

\noindent It is then easy to get what we need

\begin{equation}
L_{\mathrm{radius}}=\frac{1}{2}L_{\mathrm{boundary}}=R\log\left(\frac{L_{S}}{\pi a}\right).
\end{equation}

\noindent We now map $L_{S}\to\beta$ to get the geodesic length between
$a$ and $\beta_{H}=\frac{\beta}{2\pi}$ in the finite temperature
system:

\begin{equation}
L=R\log\left(\frac{\beta}{\pi a}\right)=R\log\left(\frac{2\beta_{H}}{a}\right),\label{eq:step5}
\end{equation}

\noindent which agrees with that obtained   through holographic duality in  \cite{Takayanagi:2017knl}.
Therefore, from the general expression (\ref{eq:general expression of L}),
as $x=x^{\prime}$, $t=t^{\prime}$, $y=a$ and $y^{\prime}=\beta_{H}$
, we have
\begin{eqnarray}
L_{\mathrm{boundary}} & = & R\log\left(\frac{2\beta_{H}}{a}\left[f\left(\frac{a}{\beta_{H}},\frac{\beta_{H}}{\beta_{H}}\right)-g\left(\frac{a}{\beta_{H}},\frac{\beta_{H}}{\beta_{H}}\right)\right]\right)\nonumber \\
 & \rightarrow & R\log\frac{2\beta_{H}}{a}.
\end{eqnarray}
\noindent We thus obtain

\begin{equation}
f\left(\frac{a}{\beta_{H}},\frac{\beta_{H}}{\beta_{H}}\right)-g\left(\frac{a}{\beta_{H}},\frac{\beta_{H}}{\beta_{H}}\right)=1.
\end{equation}


\noindent For convenience, we summarize all the constraints we have
obtained for the general expression (\ref{eq:general expression of L})
of bulk geodesic length:
\begin{eqnarray}
 &  & f\left(\frac{y}{\beta_{H}},\frac{y^{\prime}}{\beta_{H}}\right)-g\left(\frac{y}{\beta_{H}},\frac{y^{\prime}}{\beta_{H}}\right)=\frac{1}{2\beta_{H}^{2}}\left(y^{2}+y^{\prime2}\right)+\mathcal{O}\left(\frac{y^{4}}{\beta_{H}^{4}}\right),\hspace{1em}\beta_{H}\gg y,y',\label{eq:R3}\\
\nonumber \\
 &  & f\left(\frac{y}{\beta_{H}},\frac{\beta_{H}}{\beta_{H}}\right)^{2}=f\left(\frac{y}{\beta_{H}},\frac{y}{\beta_{H}}\right)=1,\label{eq:R4}\\
 &  & f\left(\frac{y}{\beta_{H}},\frac{y}{\beta_{H}}\right)-g\left(\frac{y}{\beta_{H}},\frac{y}{\beta_{H}}\right)=\frac{y^{2}}{\beta_{H}^{2}},\label{eq:R5}\\
 &  & f\left(\frac{a}{\beta_{H}},\frac{\beta_{H}}{\beta_{H}}\right)-g\left(\frac{a}{\beta_{H}},\frac{\beta_{H}}{\beta_{H}}\right)=1.\label{eq:R6}
\end{eqnarray}
\noindent From eqn. (\ref{eq:R4}) and (\ref{eq:R6}), we get

\begin{equation}
g\left(\frac{a}{\beta_{H}},\frac{\beta_{H}}{\beta_{H}}\right)=0.
\end{equation}

\noindent Since $a$ is a varying quantity, $y$ or $y'=\beta_{H}$
must be a zero of $g(y/\beta_{H},y'/\beta_{H})$ . Moreover, $g$
must be symmetric for $y$ and $y'$. So, the function form must be

\begin{equation}
g\left(\frac{y}{\beta_{H}},\frac{y'}{\beta_{H}}\right)\propto\left(1-\frac{y^{n}}{\beta_{H}^{n}}\right)^{\kappa}\left(1-\frac{y'^{n}}{\beta_{H}^{n}}\right)^{\kappa}\left(\cdots\right)
\end{equation}

\noindent On the other hand, from eqn. (\ref{eq:R4}) and (\ref{eq:R5}),
one gets

\begin{equation}
g\left(\frac{y}{\beta_{H}},\frac{y}{\beta_{H}}\right)=1-\frac{y^{2}}{\beta_{H}^{2}}
\end{equation}

\noindent It then easy to fix $n=2$ and $\kappa=1/2$ and

\begin{equation}
g\left(\frac{y}{\beta_{H}},\frac{y'}{\beta_{H}}\right)=\sqrt{\left(1-\left(\frac{y}{\beta_{H}}\right)^{2}\right)\left(1-\left(\frac{y^{\prime}}{\beta_{H}}\right)^{2}\right)}\left[1+\left(\frac{\Delta y}{\beta_{H}}\right)^{2}\left(\sigma_{1}+{\cal O}\left(\frac{y}{\beta}\right)\right)\right].\label{eq:g expansion}
\end{equation}

\noindent Similarly, from eqn. (\ref{eq:R4}), we get

\begin{equation}
f\left(\frac{y}{\beta_{H}},\frac{y'}{\beta_{H}}\right)=1+\left(\frac{\Delta y}{\beta_{H}}\right)^{2}\left(1-\frac{y^{m}}{\beta_{H}^{m}}\right)^{\delta}\left(1-\frac{y'^{m}}{\beta_{H}^{m}}\right)^{\delta}\left[\theta_{1}+{\cal O}\left(\frac{y}{\beta}\right)+\cdots\right],\label{eq:f expansion}
\end{equation}

\noindent where $m$, $\delta>0$ are some numbers. The story is not over yet. In order to match eqn. (\ref{eq:R3}),
there must be $\sigma_{1}=\theta_{1}$. When applying eqn. (\ref{eq:derive metric})
to calculate the metric, noting a limit $\Delta x,\Delta y,\Delta t\to0$
is going to be imposed after making the derivatives, it is easy to
see that terms proportional to $\left(\frac{\Delta y}{\beta_{H}}\right)^{2}$
only contribute to $g_{yy}$, but not to $g_{xx}$ and $g_{tt}$.
So, looking at eqn. (\ref{eq:general expression of L}), (\ref{eq:g expansion}) and (\ref{eq:f expansion}),
without altering the fixed metric, equivalently, we are free to
pack all the corrections into $g(y,y')$ and simply set $f(y,y')=1$.
Finally, we   substitute $f(y,y')$ and $g(y,y')$ into eqn. (\ref{eq:general expression of L}) to get the bulk geodesic length,
\begin{eqnarray}
\cosh\left(\frac{L_{\mathrm{bulk}}}{R}\right) & = &\frac{\beta_{H}^{2}}{yy^{\prime}} \left[\cosh\left(\frac{\triangle x}{\beta_{H}}\left(1-\frac{\pi\mu c}{12\beta_{H}^{2}}+\ldots\right)\right)\right.\nonumber \\
 &  & \left.-\sqrt{\left(1-\left(\frac{y}{\beta_{H}}\right)^{2}\right)\left(1-\left(\frac{y^{\prime}}{\beta_{H}}\right)^{2}\right)}\left(1+\left(\frac{\Delta y}{\beta_{H}}\right)^{2}\cdot{\cal O}\left(\frac{y}{\beta}\right)\right)\cosh\left(\frac{\triangle t}{\beta_{H}}\right)\right].
\end{eqnarray}
\noindent Using (\ref{eq:derive metric}), we obtain the deformed BTZ metric,
\begin{equation}
ds^{2}=\frac{R_{AdS}^{2}}{y^{2}}\left(-\left(1-\frac{y^{2}}{\beta_{H}^{2}}\right)dt^{2}+\left(1-\frac{\pi\mu c}{12\beta_{H}^{2}}+\ldots\right)^{2}dx^{2}+\left(1-\frac{y^{2}}{\beta_{H}^{2}}\right)^{-1}\left(1+{\cal O}\left(\frac{y}{\beta_{H}}\right)\right)dy^{2}\right).\label{eq:finial metric}
\end{equation}

\noindent Note the coordinate $x$ and $t$ here are the physical
arguments of the $T\bar{T}$ deformed CFT$_{2}$. In contrast, in the refs \cite{McGough:2016lol,Chen:2018eqk},
the authors started with the standard BTZ metric, and then the boundary coordinates in their works
are the physical arguments of the undeformed CFT. To have the deformed
CFT living on a conformally flat boundary, from this deformed BTZ metric (\ref{eq:finial metric}),  there ought to be %

\begin{equation}
\left(1-\frac{y_{c}^{2}}{\beta_{H}^{2}}\right)=\left(1-\frac{\pi\mu c}{12\beta_{H}^{2}}+\ldots\right)^{2}.\label{eq:yc}
\end{equation}

\noindent So the boundary of the dual geometry is located at $y^2_{c}=\frac{\pi c\mu}{6}+\ldots$,   which is finite. Transforming the planar coordinate to the global one by $y=R_{AdS}^{2}/r$, we get

\begin{equation}
y^2_{c}=\frac{\pi c\mu}{6}+\ldots,\hspace{5mm}\hbox{or}\hspace{5mm} r_{c}^{2}=\frac{6R_{AdS}^{4}}{\pi c\mu}+\ldots, \label{eq:yc1}
\end{equation}

\noindent which completely agrees with (\ref{eq:radial}) as conjectured
in ref. \cite{McGough:2016lol}. One of the advantages of the deformed metric (\ref{eq:finial metric}) is that it
naturally indicates that the dual geometry is a finite region $y_{c}\leq y \leq\beta_{H}$. For simplicity, we rewrite the metric by using eqn. (\ref{eq:yc1}) as
\begin{equation}
ds^{2}=\frac{R_{AdS}^{2}}{y^{2}}\left(-\left(1-\frac{y^{2}}{\beta_{H}^{2}}\right)dt^{2}+\left(1-\frac{y_{c}^{2}}{\beta_{H}^{2}}\right)dx^{2}+\left(1-\frac{y^{2}}{\beta_{H}^{2}}\right)^{-1}\left(1+{\cal O}\left(\frac{y}{\beta_{H}}\right)\right)dy^{2}\right).
\label{eq:TTMetric}
\end{equation}

It is worth noting that, in a recent development of $T\bar{T}$ deformed CFT \cite{Lewkowycz:2019xse}, the authors obtained a remarkable   all-orders result for the entanglement entropy in the Poincare disk at finite $\mu$. That nonperturbative result   might be of help to fix the identification in  (\ref{eq:yc}) nonperturbatively, and eliminate the higher order corrections in $(y/\beta)$ of the metric component $g_{yy}$ in eqn. (\ref{eq:TTMetric}).
The tricky part is that  that  nonperturbative result obtained in \cite{Lewkowycz:2019xse} from the field theory side is  $d  S^\mu_{EE}/ d\left(\Delta x\right)$, but not  the entanglement entropy  itself.
When we integrate it  to obtain  $S_{EE}^{\mu}\left(\Delta x\right)$,   the  integration constant contains information about the holographic direction $y$ and we must be very careful to explore its role.  We wish to study this question in future work.

To study the propagation speed, let us consider the induced metric
at some point  $y=y_{*}$:

\begin{equation}
\left.ds^{2}\right|_{y=y_{*}}=\frac{R_{AdS}^{2}}{y_{*}^{2}}\left(-\left(1-\frac{y_{*}^{2}}{\beta_{H}^{2}}\right)dt^{2}+\left(1-\frac{y_{c}^{2}}{\beta_{H}^{2}}\right)dx^{2}\right)=\frac{R_{AdS}^{2}}{y_{*}^{2}}\left(-dt_{*}^{2}+dx_{*}^{2}\right),
\end{equation}

\noindent where we define

\begin{equation}
dt_{*}=\left(1-\frac{y_{*}^{2}}{\beta_{H}^{2}}\right)^{\frac{1}{2}}dt,\qquad dx_{*}=\left(1-\frac{y_{c}^{2}}{\beta_{H}^{2}}\right)^{\frac{1}{2}}dx.
\end{equation}

\noindent The right and left propagation speeds of the wave are given
by

\begin{equation}
v_{\pm}=\pm\frac{dx_{*}}{dt_{*}}=\frac{\left(1-\frac{y_{c}^{2}}{\beta_{H}^{2}}\right)^{\frac{1}{2}}}{\left(1-\frac{y_{*}^{2}}{\beta_{H}^{2}}\right)^{\frac{1}{2}}},\qquad\mu\geq0.
\end{equation}

\noindent Considering an observer sitting at $y_{c}^{2}=\frac{\mu\pi c}{6}$
(or $r_{c}^{2}=\frac{6R_{AdS}^{4}}{\pi c\mu}$), when $0<y_{*}^{2}< y_{c}^{2}=\frac{\mu\pi c}{6}+\ldots.$,
the   speed is   less than the speed of light (subluminal). When
$y_{*}^{2}>y_{c}^{2}=\frac{\mu\pi c}{6}+\ldots.$, the  speed
is superluminal, which again agrees with   the results of $T\bar{T}$
deformed CFT \cite{McGough:2016lol}. More discussions can be found in \cite{Brattan:2011my,Marolf:2012dr}.

As a further check, we compute the proper energy by using the deformed
BTZ metric eqn. (\ref{eq:finial metric}). Transforming the planar coordinate to the global one, setting $R_{AdS}=1$ for simplicity, we then have

\begin{equation}
ds^{2}=-f\left(r\right)^{2}dt^{2}+\frac{dr^{2}}{f\left(r\right)^{2}}+r^{2} d\phi^{2},\label{eq:global metric}
\end{equation}

\noindent with

\begin{equation}
f\left(r\right)^{2}=r^{2}-8GM,\qquad r_{H}^{2}=8GM,\qquad\phi\sim\phi+2\pi.
\end{equation}

\noindent Since we already determined  the boundary of the geometry   at $r_c$, when calculating the  quasi-local gravitational energy, we should integrate along $r=r_c$.
Using $\frac{3}{2Gc}=R_{AdS}=1$, the proper energy  \cite{Brown:1992br,Balasubramanian:1999re,Kraus:2018xrn} is
\begin{eqnarray}
\mathcal{E} & = & EL=\frac{L}{2\pi}\int_{0}^{2\pi}d\phi\sqrt{g_{\phi\phi}}\mu^{i}\mu^{j}T_{ij}\Big|_{r=r_c}\nonumber \\
 & = & \frac{\pi r_{c}^{2}}{2G}\left[1-\sqrt{1-\frac{8GM}{r_{c}^{2}}}\right]\nonumber \\
&=& \frac{2}{\mu}\Big[ 1-\sqrt{1-2\pi\mu M} \Big].
\end{eqnarray}
\noindent where $L=\int d\phi \sqrt{g_{\phi\phi}}\,|_{r=r_c}$ is the proper size on the boundary,  and the non-vanishing components are

\begin{equation}
\mu^{t}=\frac{1}{f\left(r\right)},\qquad T_{tt}=\frac{1}{4G}\left(f\left(r\right)^{2}-\frac{1}{r}f\left(r\right)^{3}\right).
\end{equation}

\noindent This energy  completely agrees with the energy spectrum of $T\bar{T}$ deformed
CFT$_2$, given in \cite{McGough:2016lol,Kraus:2018xrn}.

\section{The dual geometry of the highly excited states in CFT$_{2}$}

For a highly excited state of CFT$_{2}$,

\begin{equation}
\left|\psi_{h}\right\rangle =\underset{z\rightarrow0}{\lim}z^{\triangle_{h}}\mathcal{O}_{h}\left(z\right)\left|0\right\rangle ,
\end{equation}

\noindent where $\triangle_{h}\sim\mathcal{O}\left(c\right)$ is the
conformal dimension of a heavy primary operator $\mathcal{O}_{h}$,
usually it is very difficult to calculate the entanglement entropy.
Using the replica trick, the entanglement entropy can be calculated
by a four point function, which consists of two heavy operators exciting
the states and two light twist operators. This four point function
can be expanded in conformal blocks with OPE. The problem is that
generically the closed form of the conformal blocks is unknown. However,
it is believed that, in order to have a gravity dual, a CFT should
have a large central charge and a sparse spectrum of light operators.
Remarkably, for such CFTs, the conformal block expansion is well approximated
by the identity block and the entanglement entropy of a finite size
system is \cite{Fitzpatrick:2014vua, Asplund:2014coa,Caputa:2014eta}

\begin{equation}
S_{EE}=\frac{c}{3}\log\left(\frac{2\beta_{\Psi}}{a}\sinh\left(\frac{\triangle x}{2\beta_{\Psi}}\right)\right),\label{eq:excited EE}
\end{equation}
with $\beta_{\Psi}=\left(\sqrt{\frac{12\triangle_{h}}{c}-1}\right)^{-1}$.
This entanglement entropy is identical to that of a finite temperature
system with $T=\beta_{\Psi}^{-1}$. As three dimensional gravity has
no propagation, the excited states of the dual CFT$_{2}$
only leads to local defects and global topologies in the bulk. This
is why the entanglement entropies of finite size system, finite temperature
system and their excited states take the same form, and their dual
metrics are connected by simple transformations. It is then easy to
understand that we really only need to consider a single representative
of CFTs with distinct topologies.

\noindent So, we can safely use the same procedure which we have
used in \cite{Wang:2018vbw}  to fix the leading behaviors of the dual geometries of CFT$_2$ at finite temperature to get
the geodesic length:

\begin{equation}
\cosh\left(\frac{L_{\mathrm{bulk}}}{R}\right)=\left[\cosh\left(\frac{\triangle x}{\beta_{\Psi}}\right)-\sqrt{\left(1-\left(\frac{y}{\beta_{\Psi}}\right)^{2}\right)\left(1-\left(\frac{y^{\prime}}{\beta_{\Psi}}\right)^{2}\right)}\left(1+\left(\frac{\Delta y}{\beta_{\Psi}}\right)^{2}\cdot{\cal O}\left(\frac{y}{\beta_{\Psi}}\right)\right)\cosh\left(\frac{\triangle t}{\beta_{\Psi}}\right)\right],
\end{equation}

\noindent and the metric

\begin{equation}
ds^{2}=\frac{R_{AdS}^{2}}{y^{2}}\left(-\left(1-\frac{y^{2}}{\beta_{\Psi}^{2}}\right)dt^{2}+\left(1-\frac{y^{2}}{\beta_{\Psi}^{2}}\right)^{-1}\left(1+{\cal O}\left(\frac{y}{\beta_{\Psi}}\right)\right)dy^{2}+dx^{2}\right)
\end{equation}

\noindent Setting $R_{AdS}=1$ for simplicity,
the metric in globe coordinate system is

\begin{equation}
ds^{2}=-\left(r^{2}+1-\frac{12\triangle_{h}}{c}\right)dt^{2}+\frac{dr^{2}}{\left(r^{2}+1-\frac{12\triangle_{h}}{c}\right)}\left(1+\ldots\right)+r^{2}d\phi^{2}.\label{eq:excited metric}
\end{equation}

\noindent When $\triangle_{h}<\frac{c}{12}$, $\beta_{\Psi}$ is imaginary,
it is a finite size system with rescaled length. The dual geometry
describes a conical defect placed in the center of the global AdS. The defect is caused by the back-reaction of a massive particle.
When $\triangle_{h}>\frac{c}{12}$, the primary state $\left|\psi_{h}\right\rangle $
approximates to a thermal state and the dual excited state in AdS
is heavy enough to form a black hole. These conclusions are in consistent
with the holographic calculations \cite{Asplund:2014coa,Caputa:2014eta}.

\section{Conclusion}

In this paper, we fixed the leading behaviors of the deformed  BTZ black hole metric from
the entanglement entropy of $T\bar{T}$ deformed CFT$_{2}$. The metric
shows explicitly that the dual region $\frac{6R_{AdS}^{4}}{c\mu}\geq r^{2}\geq r_{H}^{2}$
is finite in the bulk. We used the deformed  BTZ metric to calculate
the propagation speed and energy spectrum. Both results match perfectly
with those calculated in the $T\bar{T}$ deformed CFT$_{2}$, and
no identification is necessary. We then showed how to get the dual
geometry of highly excited states of CFT$_{2}$. The metric describes
a conical defect located in the center of the global AdS for $\triangle_{h}<\frac{c}{12}$,
or covers a BTZ black hole at temperature $T=\beta_{\Psi}^{-1}=\sqrt{\frac{12\triangle_{h}}{c}-1}$
as $\triangle_{h}>\frac{c}{12}$.

Finally, we wish to clarify two significant reasons that why our derivations may look  heavy:

\begin{enumerate}
\item The advantages of our method is to also cover the   time-like
direction, i.e. the full spacetime metric, naturally. It is very difficult
because we only know the information of the lower dimensional theories.

\item Our purpose is not only to figure out the linear order but also the singularity and event horizon of BTZ spacetime. The behaviors of black hole's singularity and event horizon can not be extracted by the leading term of the spacetime metric directly. This is why we use more results of entanglement entropies, and aim to fix more accurate leading behaviors of bulk geometries. 
\end{enumerate}

Moreover, in this paper, our aim is not to explain how bulk
geometry is emerged from entanglement entropies, or to derive the bulk
dynamics (Einstein's equation) from the boundary theory, but only to show that
the entanglement entropies of $\mathrm{CFT}_{2}$ are enough to fix
the leading behaviors of the bulk spacetime geometries.

\vspace{5mm}

\noindent {\bf Acknowledgements}
We are deeply indebt to Bo Ning  for many illuminating discussions and suggestions.  We are also very grateful to Qingyu Gan,   Sung-Soo Kim,  Zhao-Long Wang and Shuxuan Ying for very helpful discussions and suggestions. This work is supported in part by the NSFC (Grant No. 11875196, 11375121 and 11005016). 


\begin{thebibliography}{99}
\bibitem{Zamolodchikov:2004ce}    A.~B.~Zamolodchikov,   ``Expectation value of composite field T anti-T in two-dimensional quantum field theory,''   hep-th/0401146.   

\bibitem{Cavaglia:2016oda}
  A.~Cavaglia, S.~Negro, I.~M.~Szecsenyi and R.~Tateo,
  ``$T \bar{T}$-deformed 2D Quantum Field Theories,''
  JHEP {\bf 1610}, 112 (2016)
  doi:10.1007/JHEP10(2016)112
  [arXiv:1608.05534 [hep-th]].


\bibitem{Smirnov:2016lqw}
F.~A.~Smirnov and A.~B.~Zamolodchikov,   ``On space of integrable quantum field theories,''   Nucl.\ Phys.\ B {\bf 915}, 363 (2017)   doi:10.1016/j.nuclphysb.2016.12.014   [arXiv:1608.05499 [hep-th]].   



\bibitem{McGough:2016lol}
L.~McGough, M.~Mezei and H.~Verlinde,   ``Moving the CFT into the bulk with $ T\overline{T} $,''   JHEP {\bf 1804}, 010 (2018)   doi:10.1007/JHEP04(2018)010   [arXiv:1611.03470 [hep-th]].   


\bibitem{Chen:2018eqk}
B.~Chen, L.~Chen and P.~X.~Hao,   ``Entanglement Entropy in $T\overline{T}$-Deformed CFT,''   arXiv:1807.08293 [hep-th].   








\bibitem{Ryu:2006bv}
S.~Ryu and T.~Takayanagi,   ``Holographic derivation of entanglement entropy from AdS/CFT,''   Phys.\ Rev.\ Lett.\  {\bf 96}, 181602 (2006)   doi:10.1103/PhysRevLett.96.181602   [hep-th/0603001].   



\bibitem{Donnelly:2018bef}
  W.~Donnelly and V.~Shyam,
  ``Entanglement entropy and $T \overline{T}$ deformation,''
  Phys.\ Rev.\ Lett.\  {\bf 121}, 131602 (2018)
  doi:10.1103/PhysRevLett.121.131602
  [arXiv:1806.07444 [hep-th]].

\bibitem{Dubovsky:2012wk}
  S.~Dubovsky, R.~Flauger and V.~Gorbenko,
  ``Solving the Simplest Theory of Quantum Gravity,''
  JHEP {\bf 1209}, 133 (2012)
  doi:10.1007/JHEP09(2012)133
  [arXiv:1205.6805 [hep-th]].

\bibitem{Caselle:2013dra}
  M.~Caselle, D.~Fioravanti, F.~Gliozzi and R.~Tateo,
  ``Quantisation of the effective string with TBA,''
  JHEP {\bf 1307}, 071 (2013)
  doi:10.1007/JHEP07(2013)071
  [arXiv:1305.1278 [hep-th]].


\bibitem{Giveon:2017myj}
  A.~Giveon, N.~Itzhaki and D.~Kutasov,
  ``A solvable irrelevant deformation of AdS$_{3}$/CFT$_{2}$,''
  JHEP {\bf 1712}, 155 (2017)
  doi:10.1007/JHEP12(2017)155
  [arXiv:1707.05800 [hep-th]].

\bibitem{Giveon:2017nie}
  A.~Giveon, N.~Itzhaki and D.~Kutasov,
  ``$ \mathrm{T}\overline{\mathrm{T}} $ and LST,''
  JHEP {\bf 1707}, 122 (2017)
  doi:10.1007/JHEP07(2017)122
  [arXiv:1701.05576 [hep-th]].

\bibitem{Chakraborty:2018kpr}
  S.~Chakraborty, A.~Giveon, N.~Itzhaki and D.~Kutasov,
  ``Entanglement beyond AdS,''
  Nucl.\ Phys.\ B {\bf 935}, 290 (2018)
  doi:10.1016/j.nuclphysb.2018.08.011
  [arXiv:1805.06286 [hep-th]].


\bibitem{Shyam:2017znq}
  V.~Shyam,
  ``Background independent holographic dual to $T\bar{T}$ deformed CFT with large central charge in 2 dimensions,''
  JHEP {\bf 1710}, 108 (2017)
  doi:10.1007/JHEP10(2017)108
  [arXiv:1707.08118 [hep-th]].

\bibitem{Guica:2017lia}
  M.~Guica,
  ``An integrable Lorentz-breaking deformation of two-dimensional CFTs,''
  SciPost Phys.\  {\bf 5}, 048 (2018)
  doi:10.21468/SciPostPhys.5.5.048
  [arXiv:1710.08415 [hep-th]].

\bibitem{Giribet:2017imm}
  G.~Giribet,
  ``$T\bar{T}$-deformations, AdS/CFT and correlation functions,''
  JHEP {\bf 1802}, 114 (2018)
  doi:10.1007/JHEP02(2018)114
  [arXiv:1711.02716 [hep-th]].

\bibitem{Dubovsky:2018bmo}
  S.~Dubovsky, V.~Gorbenko and G.~Hernndez-Chifflet,
  ``$ T\overline{T} $ partition function from topological gravity,''
  JHEP {\bf 1809}, 158 (2018)
  doi:10.1007/JHEP09(2018)158
  [arXiv:1805.07386 [hep-th]].

\bibitem{Cottrell:2018skz}
  W.~Cottrell and A.~Hashimoto,
  ``Comments on $T \bar T$ double trace deformations and boundary conditions,''
  arXiv:1801.09708 [hep-th].


\bibitem{Bonelli:2018kik}
  G.~Bonelli, N.~Doroud and M.~Zhu,
  ``$T \bar{T}$-deformations in closed form,''
  JHEP {\bf 1806}, 149 (2018)
  doi:10.1007/JHEP06(2018)149
  [arXiv:1804.10967 [hep-th]].


\bibitem{Akhavan:2018wla}
  A.~Akhavan, M.~Alishahiha, A.~Naseh and H.~Zolfi,
  ``Complexity and Behind the Horizon Cut Off,''
  arXiv:1810.12015 [hep-th].














\bibitem{Wang:2017bym}
P.~Wang, H.~Wu and H.~Yang,   ``AdS$_3$ metric from UV/IR entanglement entropies of CFT$_2$,''   arXiv:1710.08448 [hep-th].   

\bibitem{Wang:2018vbw}
P.~Wang, H.~Wu and H.~Yang,   ``Derive three dimensional geometries from entanglement entropies of CFT$_2$,''   arXiv:1809.01355 [hep-th].   

\bibitem{Synge:1960}
J. Synge,   ``Relativity: The general theory,''    North-Holland, Amsterdam, The Netherlands, 1960.


\bibitem{Fitzpatrick:2014vua}    A.~L.~Fitzpatrick, J.~Kaplan and M.~T.~Walters,   ``Universality of Long-Distance AdS Physics from the CFT Bootstrap,''   JHEP {\bf 1408}, 145 (2014)   doi:10.1007/JHEP08(2014)145   [arXiv:1403.6829 [hep-th]].   

\bibitem{Asplund:2014coa}
C.~T.~Asplund, A.~Bernamonti, F.~Galli and T.~Hartman,   ``Holographic Entanglement Entropy from 2d CFT: Heavy States and Local Quenches,''   JHEP {\bf 1502}, 171 (2015)   doi:10.1007/JHEP02(2015)171   [arXiv:1410.1392 [hep-th]].   

\bibitem{Caputa:2014eta}    P.~Caputa, J.~Sim\'on, A.~\v Stikonas and T.~Takayanagi,   ``Quantum Entanglement of Localized Excited States at Finite Temperature,''   JHEP {\bf 1501}, 102 (2015)   doi:10.1007/JHEP01(2015)102   [arXiv:1410.2287 [hep-th]].   


\bibitem{Czech:2015qta}
B.~Czech, L.~Lamprou, S.~McCandlish and J.~Sully, ``Integral Geometry and Holography,'' JHEP \textbf{10}, 175 (2015) doi:10.1007/JHEP10(2015)175 [arXiv:1505.05515 [hep-th]]. 

\bibitem{Czech:2016xec}
B.~Czech, L.~Lamprou, S.~McCandlish, B.~Mosk and J.~Sully, ``A Stereoscopic Look into the Bulk,'' JHEP \textbf{07}, 129 (2016) doi:10.1007/JHEP07(2016)129 [arXiv:1604.03110 [hep-th]]. 

\bibitem{deBoer:2016pqk}
J.~de Boer, F.~M.~Haehl, M.~P.~Heller and R.~C.~Myers, ``Entanglement, holography and causal diamonds,'' JHEP \textbf{08}, 162 (2016) doi:10.1007/JHEP08(2016)162 [arXiv:1606.03307 [hep-th]]. 



\bibitem{Takayanagi:2011zk} T.~Takayanagi, ``Holographic Dual of BCFT,'' Phys. Rev. Lett. \textbf{107}, 101602 (2011) doi:10.1103/PhysRevLett.107.101602 [arXiv:1105.5165 [hep-th]]. 

\bibitem{Sully:2020pza} J.~Sully, M.~Van Raamsdonk and D.~Wakeham, ``BCFT entanglement entropy at large central charge and the black hole interior,'' [arXiv:2004.13088 [hep-th]]. 



\bibitem{Calabrese:2004eu}    P.~Calabrese and J.~L.~Cardy,   ``Entanglement entropy and quantum field theory,''   J.\ Stat.\ Mech.\  {\bf 0406}, P06002 (2004)   doi:10.1088/1742-5468/2004/06/P06002   [hep-th/0405152].   

\bibitem{Calabrese:2009qy}    P.~Calabrese and J.~Cardy,   ``Entanglement entropy and conformal field theory,''   J.\ Phys.\ A {\bf 42}, 504005 (2009)   doi:10.1088/1751-8113/42/50/504005   [arXiv:0905.4013 [cond-mat.stat-mech]].   

\bibitem{Rangamani:2016dms}
M.~Rangamani and T.~Takayanagi,
Lect. Notes Phys. \textbf{931}, pp.1-246 (2017)
doi:10.1007/978-3-319-52573-0
[arXiv:1609.01287 [hep-th]].


\bibitem{Takayanagi:2017knl}    K.~Umemoto and T.~Takayanagi,   ``Entanglement of purification through holographic duality,''   Nature Phys.\  {\bf 14}, no. 6, 573 (2018)   doi:10.1038/s41567-018-0075-2   [arXiv:1708.09393 [hep-th]].   

\bibitem{Lewkowycz:2019xse}
A.~Lewkowycz, J.~Liu, E.~Silverstein and G.~Torroba, ``$ T\overline{T} $ and EE, with implications for (A)dS subregion encodings,'' JHEP \textbf{04}, 152 (2020) doi:10.1007/JHEP04(2020)152 [arXiv:1909.13808 [hep-th]]. 


\bibitem{Brattan:2011my}
  D.~Brattan, J.~Camps, R.~Loganayagam and M.~Rangamani,
  ``CFT dual of the AdS Dirichlet problem : Fluid/Gravity on cut-off surfaces,''
  JHEP {\bf 1112}, 090 (2011)
  doi:10.1007/JHEP12(2011)090
  [arXiv:1106.2577 [hep-th]].

\bibitem{Marolf:2012dr}
  D.~Marolf and M.~Rangamani,
  ``Causality and the AdS Dirichlet problem,''
  JHEP {\bf 1204}, 035 (2012)
  doi:10.1007/JHEP04(2012)035
  [arXiv:1201.1233 [hep-th]].


\bibitem{Brown:1992br}
  J.~D.~Brown and J.~W.~York, Jr.,
  ``Quasilocal energy and conserved charges derived from the gravitational action,''
  Phys.\ Rev.\ D {\bf 47}, 1407 (1993)
  doi:10.1103/PhysRevD.47.1407
  [gr-qc/9209012].

\bibitem{Balasubramanian:1999re}
  V.~Balasubramanian and P.~Kraus,
  ``A Stress tensor for Anti-de Sitter gravity,''
  Commun.\ Math.\ Phys.\  {\bf 208}, 413 (1999)
  doi:10.1007/s002200050764
  [hep-th/9902121].

\bibitem{Kraus:2018xrn}
P.~Kraus, J.~Liu and D.~Marolf,   ``Cutoff AdS$_{3}$ versus the $ T\overline{T} $ deformation,''   JHEP {\bf 1807}, 027 (2018)   doi:10.1007/JHEP07(2018)027   [arXiv:1801.02714 [hep-th]].   



\end{thebibliography}
\end{document}